\documentclass[reprint, 
groupedaddress,
superscriptaddress,
longbibliography,
amsmath,amssymb,
aps,
pra,
]{revtex4-2}
\usepackage{graphicx}
\usepackage{dcolumn}
\usepackage{bm}
\usepackage{xcolor}
\usepackage{braket}
\usepackage{soul}
\usepackage{comment}
\usepackage[colorlinks = true,]{hyperref}

\begin{document}
\author{Ankit Kundu}
\affiliation{Elmore Family School of Electrical and Computer Engineering, Purdue University, West Lafayette, IN 47906, USA}
\author{Rahul Trivedi}
\email{rahul.trivedi@mpq.mpg.de}
\affiliation{Max Planck Institute of Quantum Optics, D-85741 Garching, Germany}
\author{Alisa Javadi}
\email{alisa.javadi@ou.edu}
\affiliation{Department of Electrical and Computer Engineering, University of Oklahoma, Norman, OK 73069, USA} 
\affiliation{Department of Physics and Astronomy, University of Oklahoma, Norman, OK 73069, USA} 
\author{Hadiseh Alaeian}
\email{halaeian@purdue.edu}
\affiliation{Elmore Family School of Electrical and Computer Engineering, Purdue University, West Lafayette, IN 47906, USA}
\affiliation{Department of Physics and Astronomy, Purdue University, West Lafayette, IN 47906, USA}

\date{\today}

\begin{abstract}
%The Dicke model, a paradigmatic setting to explore light-matter interactions, reveals collective phenomena like superradiance, subradiance, and quantum phase transitions. 
The realization and control of collective effects in quantum emitter ensembles have predominantly focused on small, ordered systems, leaving their extension to larger, more complex configurations as a significant challenge. Quantum photonic platforms, with their engineered Green's functions and integration of advanced solid-state quantum emitters, provide opportunities to explore new regimes of light-matter interaction beyond the scope of atomic systems.  
In this study, we examine the interaction of quantum emitters embedded within a thin dielectric layer. Our results reveal that the guided optical modes of the dielectric layer mediate extended-range interactions between emitters, enabling both total and directional superradiance in arrays spanning several wavelengths. Additionally, the extended interaction range facilitated by the dielectric layer supports Dicke superradiance in regimes where collective effects cannot be obtained in a homogeneous environment. 
This work uncovers a distinctive interplay between environmental dimensionality and collective quantum dynamics, paving the way for exploring novel many-body quantum optical phenomena in engineered photonic environments.
\end{abstract}

%\title{Long-Range Dicke Superradiance in Dielectric Slabs}
\title{Cooperative Effects in Thin Dielectric Layers: Long-Range Dicke Superradiance}
\maketitle

\section{introduction}~\label{sec: intro}
As first explored by Dicke in his seminal work in 1954, the radiative properties of an atom are profoundly influenced by the presence of neighboring atoms and their quantum states, which interact \emph{collectively} through shared electromagnetic fields~\cite{Dicke1954}. A hallmark of these collective effects is \emph{superradiance}, a phenomenon that arises within an ensemble of excited emitters~\cite{Gross1982}. Unlike the standard exponential decay of independent atoms, superradiance manifests as an initial surge in emission intensity. This burst originates from the synchronization of atomic dipoles during the emission process, leading to a coherent mechanism that accelerates the emission rate. To date, the superradiant effect, aka the superradiance burst, has been observed in several experiments, including emitter ensembles in free space~\cite{Skribanowitz1973, Gross1976, Gross1979, Hettich2002, Ferioli2021, Trebbia2022, Richter2023, Ferioli2023, Lange2024, Ferioli2024} and within optical cavities~\cite{Bohnet2012, Norcia2016}.

While Dicke's original work focused on emitter ensembles in zero dimension, subsequent theoretical and experimental studies have examined collective effects in extended systems such as one-dimensional (1D) waveguides and nanofibers~\cite{Goban2015, Solano2017, Sinha2020, Asaoka2022, Pennetta2022, Skljarow2022, Pak2022, Lechner2023, Tiranov2023, Stryzhenko2024, Liedl2024}. Moreover, innovations in experimental techniques, including optical lattices and optical tweezers, now allow the precise creation, manipulation, and detection of highly ordered atomic arrays in nearly arbitrary configurations~\cite{Tao2024, Gyger2024}. These advances have spurred significant interest in exploring collective effects in two-dimensional (2D) systems, the spatial configuration of emitters, and its influence on collective phenomena~\cite{Masson2022, Rubies-Bigorda2023, Masson2024, Wenqi2024, Gjonbalaj2024}. For example, Masson et al. demonstrated that the geometry of emitter placement significantly alters the superradiance phenomenon in atomic ensembles and provided estimates for the critical distances between emitters as a function of their spatial arrangement in free space~\cite{Masson2022}. Furthermore, Robicheaux investigated the angular dependence of superradiant emissions in 1D and 2D arrays. Those findings suggest that directional superradiance is observable from atomic arrays positioned at intervals exceeding the conventional $\lambda/2\pi$~\cite{Robicheaux2021} length scale. 
%%%%%%%%%%%%%%%%%%%%%%%%%%%%
\begin{figure}[h]
    \centering
    \includegraphics[width=1\linewidth]{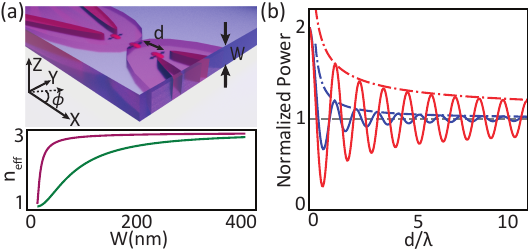}
    \caption{\textbf{Schematics.} (a) Top: Schematic representing the arrangement of quantum emitters, separated by $d$, in a dielectric layer showing spontaneous emission and particular orientations with super-radiant emission. Bottom: The effective refractive index ($n_\textrm{eff}$) of the transverse electric and transverse magnetic modes of a layer with $n = 3.5$,  as a function of layer thickness $W$, depicted in purple and green, respectively. (b) The total power radiated by two emitters, with the free-space wavelength of $\lambda_0 = 980$~nm, as a function of their separation in a homogeneous dielectric (blue) and a 200 nm-thick layer (red). The blue and red dashed-dotted lines show the asymptotic behavior of $r^{-1}$ and $r^{-0.5}$ for the homogeneous medium and thin layer, respectively. Atomic separation is normalized to the wavelength in dielectric, $\lambda$.}
    \label{fig: schematics}
\end{figure}
%%%%%%%%%%%%%%%%%%%%%%%%%%%%%%
%%%%%%%%%%%%%%%%%%%%%%%%%%%%%%%%%%%%%%%%%%%%%
\begin{figure*}[t]
    \centering
    \includegraphics[width=0.78\linewidth]{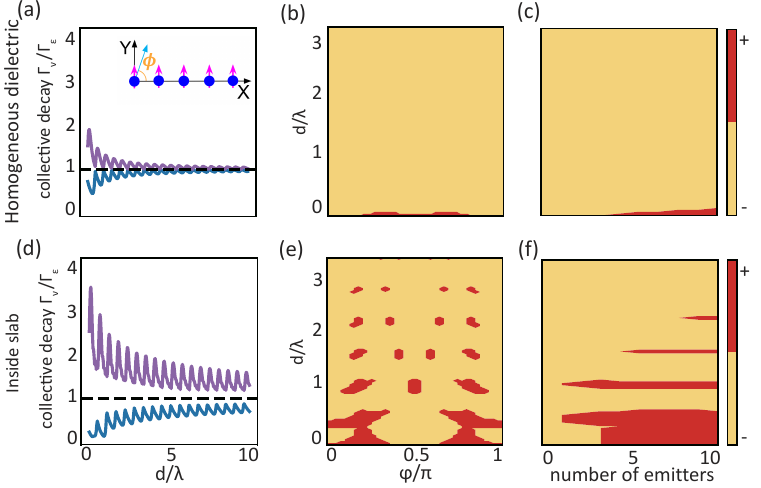}
    \caption{\textbf{Total and directional superradiance in a 1D array of Y-polarized emitters.} The collective decay rate of five emitters in a (a) homogeneous dielectric and (d) dielectric slab. The purple curve corresponds to the superradiant mode, while the blue curve corresponds to the subradiant one. Directional superradiance of a five-emitter 1D array in (b) a homogeneous dielectric and (e) dielectric slab, as a function of atomic separation ($d/\lambda$) and the in-plane angle $\phi$. The areas colored in red correspond to the regions where superradiance can be observed. The emergence of Dicke superradiance as a function of the number of emitters in a 1D array and their separations at $\phi=0.22\pi$ inside a (c) homogeneous dielectric medium and (f) slab. In all calculations $n = 3.5$ and $W = 200$ nm. }
    \label{fig: 1D-Ypol}
\end{figure*}
%%%%%%%%%%%%%%%%%%%%%%%%%%%%%%%%%%%%%%%%%%%

Here, we study the emergence of collective effects and their modifications in an array of quantum emitters embedded in a thin dielectric layer. This study is motivated by recent advances in solid-state quantum emitters, such as quantum dots and rare-earth ions in a solid-state host, which have shown highly coherent photon emission and optical linewidths that are limited by radiation~\cite{Zhai2020}. Unlike prior works, our platform addresses the role of the dimensionality of the surrounding medium. 
Our results reveal an increase in the interaction range between embedded emitters compared to that of a homogeneous medium, which is particularly significant for applications involving solid-state quantum emitters. Further, indicated by calculations and verified by numerical simulations, Dicke superradiance is robust to disorder in emitter positions. These findings contribute to a broader understanding of light-matter interactions in complex quantum systems~\cite{Manzoni2017}. Further, because of the modified interaction scaling, this platform allows us to study dissipative physics in the presence of long-range interactions. In particular, this platform could be an experimental testbed to investigate phenomena such as superluminal transport mediated by long-range interactions~\cite{foss2015nearly, tran2020hierarchy, tran2021optimal} as well as driven dissipative phases of long-range models~\cite{Schuckert2025}.
\section{Results}~\label{sec: results}
Figure~\ref{fig: schematics}(a) top shows the schematics of the problem, i.e., an (ensemble) array of quantum emitters with an (average) interatomic separation of $d$, in a dielectric layer with width $W$. For most solid-state emitters, the host is often a material with a relatively large refractive index $n$ capable of supporting several optical modes. Figure~\ref{fig: schematics}(a) bottom depicts the effective refractive index ($n_\textrm{eff}$) of the two lowest-order modes -- transverse electric (TE) and transverse magnetic (TM) -- shown in purple and green, respectively, for a free-standing dielectric layer with $n = 3.5$ as a function of $W$~\footnote{While the nomenclatures are similar to the waveguide modes, unlike waveguides', slab modes have an additional index corresponding to their variations in the azimuthal direction $\phi$.}. The modes depicted here correspond to azimuthally symmetric $\textrm{TE}_0/\textrm{TM}_0$, where the subscript has been dropped for brevity. Although in thinner layers, the TE and TM modes exhibit distinct effective indices, as $W$ increases, they asymptotically converge to the dielectric refractive index, consistent with a homogeneous medium.
 
By integrating out photonic modes, a spin model can be derived that describes photon-mediated atom-atom interactions using the photonic Green's function~\cite{Dung2002, Ruostekoski2023, Alaeian2024}. In the Markovian limit, the joint density matrix of emitters $\hat{\rho}$ evolves as ($\hbar = 1$) 
\begin{equation}~\label{eq: density_matrix}
    \dot{\hat{\rho}} = -i[\hat{H}_\textrm{A} + \hat{H}_\textrm{F} + \hat{H}_\textrm{int},  \hat{\rho}] + \mathcal{D}(\hat{\rho})\,,
\end{equation}
where $\hat{H}_\textrm{A}$ describes the Hamiltonian of the free atom and $\hat{H}_\textrm{F}$ corresponds to the external drive, and the interaction Hamiltonian $\hat{H}_\textrm{int}$ reads as
\begin{equation}~\label{eq: atom-atom interaction}
    \hat{H}_\textrm{int}  = \sum_{n \ne m} \frac{J_{mn}}{2} \left(\hat{\sigma}_{eg}^{(m)} \hat{\sigma}_{ge}^{(n)}  + \hat{\sigma}_{eg}^{(n)} \hat{\sigma}_{ge}^{(m)}\right) \,.
\end{equation}
Finally, the Lindblad terms, describing the dissipative dynamics, have the following form  
\begin{equation}~\label{eq: collective Lindblad operator}
    \mathcal{D}(\hat{\rho}) = \sum_{m,n} \frac{\Gamma_{mn}}{2} \left(2\hat{\sigma}_{ge}^{(m)} \hat{\rho} \hat{\sigma}_{eg}^{(n)} - \{\hat{\sigma}_{eg}^{(n)} \hat{\sigma}_{ge}^{(m)} , \hat{\rho}\}\right)\,.
\end{equation}
In the above equations, $\hat{\sigma}_{eg}^{(m)} = (\ket{e}\bra{g})_m$ is the raising operator and $\hat{\sigma}_{ge}^{(m)} = (\ket{g}\bra{e})_m$ is the lowering operator of the $m^\textrm{th}$ atom. Furthermore, the coherent $J_{mn}$ and dissipative $\Gamma_{mn}$ couplings are
\begin{equation}~\label{eq: pair int}
    J_{mn} = -3\pi \Gamma_0 ~  \mathrm{Re}(\mathbf{G}_\textrm{E}(r_m,r_n))  \, ,
\end{equation}
and 
\begin{equation}~\label{eq: pair decay}
    \Gamma_{mn} = 6\pi \Gamma_0 ~ \mathrm{Im}(\mathbf{G}_\textrm{E}(r_m,r_n))  \, ,
\end{equation}
where $\mathbf{G}_\textrm{E}(r_m,r_n)$ is the photonic Green's function, which depends on the frequency, the position of the emitters ($r_m,r_n$) and the macroscopic properties of the surrounding medium, and $\Gamma_0$ is the free-space decay rate of an individual atom.

In a homogeneous environment, the spherical symmetry of the environment leads to a Green's function decaying as \(r^{-1}\). In contrast, the reduced cylindrical symmetry of the slab geometry results in a decay of \(r^{-0.5}\). This slower decay indicates that the photon-mediated atom-atom interactions in Eqs.~\eqref{eq: pair int} and \eqref{eq: pair decay} extend to larger \(d\).
The extended-range interaction is evident in the radiated power of the two dipoles. Figure~\ref{fig: schematics}(b) compares the total radiated power of two emitters in a homogeneous medium and a layer with \(W = 200 \, \text{nm}\) and $n = 3.5$ as a function of their separation. The radiated power is normalized to twice that of a single emitter, indicated by the horizontal black dashed line. Deviations from this reference line signify interactions between emitters. Importantly, the emitters in the dielectric layer exhibit significant interactions for much longer $d$ than those in a homogeneous medium (red line vs. blue line). Note that the emitter spacing is normalized to the wavelength in the dielectric, i.e. $\lambda = \lambda_0/n$.  

Furthermore, the envelope of the radiated power in a homogeneous dielectric follows an asymptotic behavior as \(d^{-1}\), the blue dashed-dotted line in Fig.~\ref{fig: schematics}(b). In contrast, in a dielectric layer, the asymptotic behavior is properly captured by \(d^{-0.5}\), the red dashed-dotted line, consistent with the corresponding far-field drop-off of the slab's Green's function (cf. Appendix~\ref{app: slab GF} for the analytical full form of the Green's function). Remarkably, the interaction range in the slab extends over several wavelengths, significantly exceeding that in a homogeneous medium.

To investigate interaction effects in emitter ensembles we calculate collective decay rates, $\Gamma_\nu$, i.e. the eigenvalues of the dissipative interaction matrix $[\Gamma_{ij}]$, with $\Gamma_{ij}$ determined by $\mathbf{G}_\textrm{E}(r_m,r_n)$ as in Eq.~\eqref{eq: pair decay}~\cite{Masson2022}. More information about collective decays and their corresponding jump operators can be found in Appendix~\ref{app: SR criterion}. 
Figure~\ref{fig: 1D-Ypol}(a) and (d) present collective decay rates for a 1D array of five emitters along the X-axis and polarized along the Y-axis,  as a function of their normalized separation (\(d / \lambda\)) in a homogeneous medium and a thin layer, respectively. (cf. Fig.~\ref{fig: 1D-Zpol} for similar studies on Z-polarized emitters.)
%%%%%%%%%%%%%%%%%%%%%%%%%%%%%%%%%%%%%%%%%%%%%%%
\begin{figure}
    \centering
    \includegraphics[width=1\linewidth]{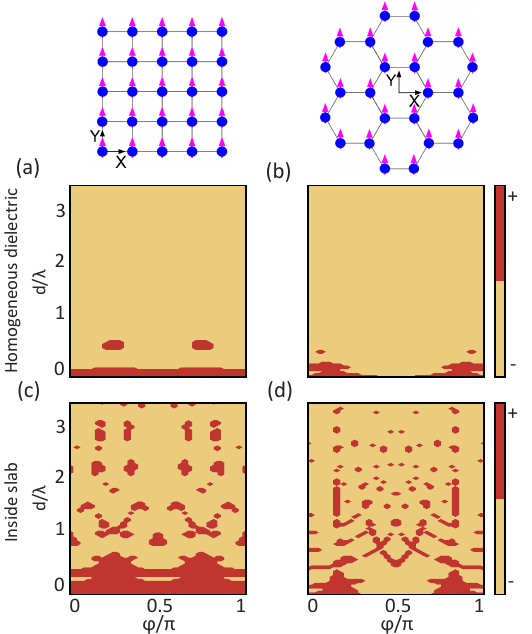}
    \caption{\textbf{Geometry effect on the total and directional superradiance in 2D arrays of Y-polarized emitters.} Directional superradiance for Y-polarized dipoles arranged in a 2D geometry inside a homogeneous dielectric ((a) and (b)) and a thin layer ((c) and (d)), delimited by red regions. Emitters in (a) and (c) are arranged on a square lattice with 25 sites, while the emitters in (b) and (d) are arranged on a hexagonal lattice with 24 sites, as depicted by small schematics on the first row. Regardless of the lattice geometry, the attainable directional superradiance regions are extended in a thin layer compared to a homogeneous medium. In all calculations $n = 3.5$ and $W = 200$ nm. Note that the hexagonal tiling will cover more area than the square one for the same lattice spacing.}
    \label{fig: 2D-Ypol}
\end{figure}
%%%%%%%%%%%%%%%%%%%%%%%%%%%%%%%%%%%%%%%%%%%%%%%

Collective decay rates are normalized to the individual emitter's in each case ($\Gamma_\epsilon$), indicated by the horizontal black dashed line. For brevity, only the highest and lowest decay rates are shown, in purple and blue, respectively. In the homogeneous dielectric, the normalized collective decay rates converge to unity at \(d \approx \lambda\). In contrast, in the slab, collective effects persist for inter-emitter separations up to \(d \approx 10\lambda\). This behavior starkly contrasts with most studies in free space, where collective effects are mainly attainable in sub-wavelength arrays. As delineated in Appendix~\ref{app: SR criterion}, a homogeneous medium imposes an upper bound on the emitter spacing \(d\) in a one-dimensional (1D) array for superradiance to occur. This restriction is lifted when considering 1D arrays within a dielectric layer, where superradiance can be expected regardless of \(d\), provided that the number of emitters is sufficiently increased (\(d \propto \frac{\ln N}{\Gamma^2}\)).

In addition to collective decay rates, the instantaneous \emph{total} photon emission rate $\gamma(t)$, as in Eq.~\eqref{eq: instantaneous decay}, is a proxy for collective effects. 
\begin{equation}~\label{eq: instantaneous decay}
    \gamma(t) = \sum_n \left[\Gamma_0 \braket{\hat{\sigma}_{eg}^{(n)} \hat{\sigma}_{ge}^{(n)}}(t) + \sum_{m \ne n} \Gamma_{mn} \braket{\hat{\sigma}_{eg}^{(m)} \hat{\sigma}_{ge}^{(n)}}(t)\right]\, ,
\end{equation}
In particular, an increase in the early-time instantaneous photon emission rate, i.e. $\dot{\gamma}(0) \ge 0$, signifies photon bunching, a hallmark of Dicke superradiance~\cite{Robicheaux2021, Masson2022}. Moreover, by studying the \emph{directional} instantaneous emission rate $\gamma(0,\phi)$ rather than the total emission, this criterion can be applied to identify the onset of superradiance in various directions $\phi$, i.e., the angle with the array axis, as depicted in Fig.~\ref{fig: schematics}(a). (cf. Appendix~\ref{app: SR criterion} for more information.)

Figures~\ref{fig: 1D-Ypol}(b) and (e) compare the emergence of directional superradiance in a 1D array as a function of $d/\lambda$, in a homogeneous dielectric and a thin layer, respectively. The red (yellow) regions indicate where \(\dot{\gamma}(0, \phi) \geq 0\) (\(\leq 0\)), delimiting the regions where superradiance can (not) be observed. As can be seen in the homogeneous dielectric, superradiance occurs only for very closely spaced arrays in limited regions. However, in the thin layer, the superradiance regions extend to far-separated emitters in several directions.  
To examine the effect of system size on collective effects, Fig.~\ref{fig: 1D-Ypol}(c) and (f) show the emergence of directional superradiance at \(\phi = 0.22\pi\) as a function of the number of emitters and \(d\). In a homogeneous dielectric, superradiance manifests itself only in large arrays with small separations. In contrast, in the thin layer, collective behavior emerges even for smaller arrays and persists for far-separated emitters, as evidenced by multiple red regions at larger \(d\), consistent with the scaling argument in Appendix~\ref{app: SR criterion}.

To investigate the role of array dimensions and geometry on collective effects, particularly directional superradiance, in Fig.~\ref{fig: 2D-Ypol} we study the directional superradiance for Y-polarized emitters arranged in 2D square and hexagonal arrays as a function of emitter separation and \(\phi\) (cf. Fig.~\ref{fig: 2D-Zpol} for similar studies on Z-polarized emitters). As in the 1D case, the sign of \(\dot{\gamma}(0)\) has been used to delimit the emergence of superradiance.  
In Figs.~\ref{fig: 2D-Ypol}(a) and (c), the red regions denote the directions in the plane where \(\dot{\gamma}(0, \phi) \geq 0\) for 25 emitters arranged in a square lattice, in a homogeneous dielectric and a 200~nm-thick layer, respectively vs. $d/\lambda$. Similar to the 1D case, slab-embedded arrays exhibit collective effects at significantly farther separations and in more directions.  

For the homogeneous case, collective effects are limited to small regions and occur only at short distances \( d \). In contrast, within the slab, these effects not only span larger distances and encompass additional regions but also display unique directional characteristics. These differences arise from geometry-dependent interference patterns that influence the emergence of collective effects in 2D lattices. Following the same scaling arguments for 1D arrays and as detailed in the Appendix~\ref{app: SR criterion}, the prolonged range of the slab's Green's function leads to the emergence of Dicke superradiance in farther separated emitters (dilute ensembles) compared to a homogeneous surrounding.
%%%%%%%%%%%%%%%%%%%%%%%%%%%%%%%%%%%%%%%%%%%%%%%
\begin{figure}
    \centering
    \includegraphics[width=\linewidth]{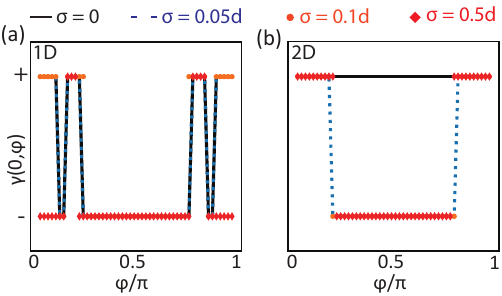}
    \caption{\textbf{Disorder effects on 1D and 2D Y-polarized emitters within a dielectric layer.} Directional superradiance 25 emitters in (a) 1D linear and (b) 2D square arrangement, with $d = 0.54 \lambda$. The black solid line, blue dashed line, orange dots, and red diamonds show the results for different disorders of $\sigma=0,~ 0.05d,~ 0.1d$, and $0.5d$, respectively. }
    \label{fig: Noise_effects_in_Ypol}
\end{figure}
%%%%%%%%%%%%%%%%%%%%%%%%%%%%%%%%%%%%%%%%%%%%%%%

To evaluate the robustness of collective effects under noise within the new interaction scaling, we investigate the impact of position disorder on directional superradiance. Noise is simulated by displacing emitters from their lattice positions by random values in the XY-plane, where the displacements are sampled from a normal distribution with zero mean and varying standard deviations ($\sigma$). The influence of disorder on directional superradiance is quantified by calculating the statistical average of $\dot{\gamma}(0,\phi)$ in multiple random configurations. Figures~\ref{fig: Noise_effects_in_Ypol}(a) and (b) illustrate the effects of noise on $\dot{\gamma}(0,\phi)$ for 25 Y-polarized emitters arranged in 1D and 2D square arrays, respectively, for various levels of disorder, $\sigma = (0,~ 0.05,~ 0.1,~ 0.5)d$. Here, $d = 0.54\lambda$, which corresponds to the approximate maximum separation at which the Y-polarized square lattice exhibits superradiance across all $\phi$, as shown in Fig.~\ref{fig: 2D-Ypol}(c). 
This analysis reveals that, despite the large disorder, there are still regions where superradiance can be obtained, as is evident in $\dot{\gamma}(0, \phi) \ge 0$ (cf. Fig.~\ref{fig: Noise effects in Zpol} in Appendix~\ref{app: z-dipoles} for noise effects on 1D and 2D arrays of Z-polarized emitters.).

\section{Discussion}~\label{sec: dicussion}
This work presents the first study of collective effects in one- and two-dimensional arrays of quantum emitters embedded within a dielectric slab. By examining the eigenvalues of the collective decoherence matrix, we show that collective phenomena persist over extended separations, surpassing the typical decay range observed in homogeneous environments. This extension arises from the modified Green’s function in slab geometries. Furthermore, we analyze the emergence of superradiance by studying the decay rate derivative, $\dot{\gamma}(0)$, for fully inverted emitter arrays. Our results demonstrate the robustness of collective effects at greater distances in reduced-dimensional systems and the emergence of pronounced directional behaviors compared to homogeneous media.

Solid-state quantum emitters, such as quantum dots, provide an ideal platform to explore these effects. Advances in achieving Fourier-limited emitters with enhanced coherence and reduced spectral wandering~\cite{Zhai2020, Zhai2022}, together with the fabrication of ordered arrays, highlight their potential. 
Future research could extend this framework to atomic arrays driven in the Heitler regime~\cite{Heitler1984, Hoffges1997, Nguyen2011, Matthiesen2012}, where the driving field phase governs the scattered photon phase. This approach could enable the states of the emitters to be entangled over larger distances. Furthermore, emerging capabilities in solid-state systems with long-range dipolar interactions open opportunities to study novel phases~\cite{Cai2013, Bilgin2024} and investigate driven-dissipative dynamics in programmable long-range interaction models~\cite{Davis2023, Block2024}.

\section*{Acknowledgment}
\noindent
HA acknowledges the insightful discussions with Alejandro Gonz\'alez-Tudela. 
Research is supported as part of the QuPIDC, an Energy Frontier Research Center, funded by the US Department of Energy (DOE), Office of Science, Basic Energy Sciences (BES), under the award number DE-SC0025620 (RT, AJ, and HA). This research was supported in part by grant number NSF PHY-2309135 to the Kavli Institute for Theoretical Physics (RT and HA). \\

\noindent
\textbf{Author contributions:} AJ, RT, and HA conceived the project, and performed the theoretical studies, while AK helped with some numerical simulations. RT, AJ, and HA wrote the manuscript and all authors contributed to the discussion of the results. 

\bibliography{ref}

\clearpage
\newpage
\appendix

\section{Dyadic Green's function of an electric dipole in stratified media}~\label{app: slab GF}
To find the Green's function of the desired problem, we need to find the Green's function of an electric dipole in a slab. The formalism presented here follows the general approach developed for stratified media~\cite{Danz2002, Novotny2012}. Here we are interested in the simplified case of one slab containing the quantum emitter, as in Fig.~\ref{fig: schematics}, given as
\begin{equation}
    \left(\nabla^2 + k_j^2\right) G(r,r') = \delta(x-x') \delta(y-y') \delta(z - z') \, ,
\end{equation}
where $k_j = \omega \sqrt{\mu_0 \epsilon_0 \epsilon_j}$ is the wavenumber of the $j^\textrm{th}$ layer with the relative permittivity of $\epsilon_j$.

Due to the translational symmetry of the problem, the polarizations of the electromagnetic field are preserved quantities. Therefore, we decompose the Green's function to $\mathbf{G}^{s(p)}_{0(\textrm{ref})}(r,r')$, where the superscript $s,p$ refers to two polarizations. In addition, subscripts $(0,\textrm{ref})$ refer to the Green's functions in the homogeneous medium and to the part due to the successive reflections from the slab interfaces. 

The dyadic Green's function of the homogeneous medium reads as follows.
\begin{widetext}
    \begin{equation*}
    \mathbf{G}_0(r,r') = \frac{i}{8\pi^2 k_2^2} \int\int_{-\infty}^{+\infty} dk_x dk_y \frac{e^{i(k_x (x-x') + k_y (y-y') + k_{z_2} |z-z'|)}}{k_{z_2}}
    \begin{bmatrix}
        k_2^2 - k_x^2 & -k_x k_y & \mp k_x k_{z_2}\\
        -k_x k_y & k_2^2 - k_y^2 & \mp k_y k_{z_2}\\
        \mp k_x k_{z_2} & \mp k_y k_{z_2} & k_2^2 - k_{z_2}^2
    \end{bmatrix}\, ,
\end{equation*}
\end{widetext}
After a bit of straightforward vector algebra, we can show that the polarization-decomposed parts are
\begin{widetext}
    \begin{equation*}
    \mathbf{G}_0^s(r,r') = \frac{i}{8\pi^2} \int\int_{-\infty}^{+\infty} dk_x dk_y \frac{e^{i(k_x (x-x') + k_y (y - y') + k_{z_2} |z-z'|)}}{k_{z_2} ~ k_\rho^2}
    \begin{bmatrix}
        k_y^2 & -k_x k_y &  0\\
        -k_x k_y & k_x^2 &  0\\
        0 &  0 & 0
    \end{bmatrix}\, ,
\end{equation*}
\end{widetext}
and
\begin{widetext}
    \begin{equation*}
    \mathbf{G}_0^p(r,r') = \frac{i}{8\pi^2} \int\int_{-\infty}^{+\infty} dk_x dk_y \frac{e^{i(k_x (x-x')+ k_y (y-y') +k_{z_2} |z-z'|)}}{k_2^2 ~ k_\rho^2}
    \begin{bmatrix}
        k_x^2 k_{z_2} & k_x k_y k_{z_2} &  \mp k_x k_\rho^2\\
        k_x k_y k_{z_2} & k_y^2 k_{z_2} &  \mp k_y k_\rho^2\\
        \mp k_x k_\rho^2 &  \mp k_y k_\rho^2 & \frac{k_\rho^4}{k_{z_2}}
    \end{bmatrix}\, ,
\end{equation*}
\end{widetext}
where $k_\rho = \sqrt{k_x^2 + k_y^2}$\, is the in-plane transverse momentum, in the direction perpendicular to the layers.

With $r_{1,2}^s(k_x,k_y)$ and $r_{1,2}^p(k_x,k_y)$ being the Fresnel reflection coefficients for $s$ and $p$-polarization at the $1^\textrm{st}$ and $2^\textrm{nd}$ interfaces, respectively, we can explicitly write the reflected Green functions as 

\begin{widetext}
\begin{align}
   & \mathbf{G}_\textrm{ref}^s(r,r') = \frac{i}{8\pi^2} \int\int_{-\infty}^{+\infty} dk_x dk_y \frac{e^{i(k_x (x-x')+ k_y (y - y') + k_{z_2} |z-z'|)}}{k_{z_2} ~ k_\rho^2}
    \begin{bmatrix}
        k_y^2 & -k_x k_y &  0\\
        -k_x k_y & k_x^2 &  0\\
        0 &  0 & 0
    \end{bmatrix} \times R_s^{(x,y,z)}(k_x,k_y)\, .
\end{align}
\end{widetext}
where 
\begin{widetext}
\begin{align}
        R_s(k_x,k_y)  = & \Gamma_s(k_x,k_y) ~ e^{ik_{z_2,}(z - z_1)} \left( r_1^s(k_x , k_y) ~ e^{ik_{z_2} l_1} + r_1^s(k_x,k_y) r_2^s(k_x,k_y) ~ e^{ik_{z_2}(l_1 + 2l_2)}\right) + \\ \nonumber
        & \Gamma_s(k_x,k_y) ~ e^{ik_{z_2}(z_2 - z)} \left(r_2^s(k_x , k_y) ~ e^{ik_{z_2} l_2} + r_1^s(k_x,k_y) r_2^s(k_x,k_y) ~ e^{ik_{z_2}(2l_1 + l_2)}\right)\, ,
        \\
        \Gamma_s(k_x,k_y) = & \frac{1}{1 - e^{i2k_{z_2}l} r_1^s(k_x , k_y) r_2^s(k_x,k_y)}\, .
\end{align}
\end{widetext}

Similarly, for the $p$-polarization with $r_{1,2}^p$, as the reflection coefficients from the first and second interfaces, we have
\begin{widetext}
\begin{align}
   & \mathbf{G}_\textrm{ref}^p(r,r') = \frac{i}{8\pi^2} \int\int_{-\infty}^{+\infty} dk_x dk_y \frac{e^{i(k_x (x-x') + k_y (y - y') + k_{z_2} |z-z'|)}}{k_2^2 ~ k_\rho^2}
    \begin{bmatrix}
        k_x^2 k_{z_2} & k_x k_y k_{z_2} &  \mp k_x k_\rho^2\\
        k_x k_y k_{z_2} & k_y^2 k_{z_2} &  \mp k_y k_\rho^2\\
        \mp k_x k_\rho^2 &  \mp k_y k_\rho^2 & \frac{k_\rho^4}{k_{z_2}}
    \end{bmatrix} \times R_p^{(x,y,z)}(k_x,k_y)\, ,
\end{align}
\end{widetext}
where 
\begin{widetext}
\begin{align}
    R_s(k_x,k_y) = & \Gamma_p(k_x,k_y) ~ e^{ik_{z_2}(z - z_1)} \left( - r_1^p(k_x , k_y) ~ e^{ik_{z_2} l_1} - r_1^p(k_x,k_y) r_2^p(k_x,k_y) ~ e^{ik_{z_2}(l_1 + 2l_2)}\right) + \\ \nonumber
    & \Gamma_p(k_x,k_y) ~ e^{ik_{z_2}(z_2 - z)} \left(~ ~ r_2^p(k_x , k_y) ~ e^{ik_{z_2} l_2} + r_1^p(k_x,k_y) r_2^p(k_x,k_y) ~ e^{ik_{z_2}(2l_1 + l_2)}\right)\, , 
    \\
    \Gamma_p(k_x,k_y) = & \frac{1}{1 - e^{i2k_{z_2}l} r_1^p(k_x , k_y) r_2^p(k_x,k_y)}\, .
\end{align}
\end{widetext}

As an aside, note that for both polarizations the poles of $\Gamma_{p,s}$ correspond to the slab modes. With the Green's functions at hand, we are now well-positioned to address the interacting ensemble problem described by the generalized spin model in Eq.~\eqref{eq: density_matrix}.

\section{Collective Decay Rates and the Emergence of Directional Superradiant in an Ensmeble}~\label{app: SR criterion}
\emph{Collective decay --} In Eq.~\eqref{eq: density_matrix} describing the spin model, while the unitary dynamics is determined via the first term on the right-hand side, the second term describes collective dissipative dynamics as detailed in Eq.~\eqref{eq: collective Lindblad operator}. We define the decay matrix $[\Gamma_{ij}]$, where the interaction-induced dissipation rate, $\Gamma_{ij}$ between $(i,j)$ pair, is determined via Eq.\eqref{eq: pair decay}. By diagonalizing this matrix, the dissipative dynamics can be re-written in the common form of
\begin{equation}~\label{eq:collective Lindblad operator2}
    \mathcal{D}(\hat{\mathcal{O}}) = \sum_\nu \frac{\Gamma_\nu}{2} \left(2\hat{L}_\nu \hat{\mathcal{O}} \hat{L}_\nu^\dagger - \{\hat{L}_\nu^\dagger \hat{L}_\nu , \hat{\mathcal{O}}\}\right)\,.
\end{equation}
Here, ${L_\nu}$ denotes the corresponding jump operators that describe the decay of the excited atoms to the ground state by emitting a photon, collectively with the rate of ${\Gamma_\nu}$.
In our analysis, we exclusively used ${\Gamma_\nu}$ as a proxy to quantify the interaction strength and range (see Figs.~\ref{fig: 1D-Ypol}(a) and (d)). However, this information can also be utilized to calculate $g^{(2)}(0)$ for any initial states, enabling the evaluation of Dicke superradiance emergence~\cite{Masson2022}.

\emph{Dimensionlaity scaling and the system-size effect -- }As discussed in the main text, superradiance means that the emission of the first facilitates the emission of the second, i.e. $g^{(2)} \ge 1$. For all atoms prepared at the excited state initially, this condition can be re-written in terms of dissipation matrix $[\Gamma_{ij}]$ entries as 
\begin{equation}~\label{eq: SR cond}
    \sum_{m \ne n}^N \Gamma_{mn}^2 \ge N \Gamma_1 \, ,
\end{equation}
where $N$ is the total number of atoms in the ensemble and $\Gamma_1$ is the individual atom decay rate. In what follows, we will provide some general arguments about the emergence of superradiance in interacting 1D and 2D ensembles when the interaction is mediated via homogeneous $\mathbf{G}_h(r,r')$ and slab $\mathbf{G}_s(r,r')$ Green's functions. Since we want to derive general scaling in this section without getting involved in too many details, we use the following dependencies of the dissipative matrix entries based on the Green's functions far-field behavior as 
\begin{equation}
    \Gamma_{mn} \propto R_{mn}^{-\alpha} ,
\end{equation}
where $R_{mn} = |\vec{r}_m - \vec{r}_n|$ is the separation between $(m,n)^\textrm{th}$ atoms and $\alpha$ = 0.5 and 1, for the slab and homogeneous case, respectively.

If atoms are arranged in a 1D array with period $d$ we have
\begin{widetext} 
\begin{align}
    \sum_{m \ne n}^N \Gamma_{mn}^2 & = \frac{1}{d^{2\alpha}} \sum_{m \ne n}^N \frac{1}{|m - n|^{2\alpha}} \approx \frac{N}{d^{2\alpha}} \sum_{n = 1}^N \frac{1}{n^{2\alpha}} = 
    \begin{cases}
        \frac{N}{d^2}  & \alpha = 1 \\
        \frac{N ~ \textrm{ln}N}{d}  & \alpha = 0.5
    \end{cases}\, .
\end{align}
\end{widetext}

From Eq.~\eqref{eq: SR cond} we get the following condition for obtaining the superradiance.  

\begin{align}
    d_\textrm{min} \propto \begin{cases}
        \frac{1}{\Gamma_1} & \alpha = 1 \\
        \frac{\textrm{ln} N}{\Gamma_1^2} & \alpha = 0.5
    \end{cases}\, .
\end{align}
Therefore, for the homogeneous case, i.e. $\alpha = 1$, there is a maximum separation limit, while for the slab case, i.e. $\alpha = 0.5$, the superradiance can always be obtained if the system size, $N$, increases.  

Following similar analyses for atoms arranged in a 2D lattice with period $d$, we arrive at

\begin{align}
    d_\textrm{min} \propto \begin{cases}
        \frac{\sqrt{\textrm{ln} N}}{\Gamma_1} & \alpha = 1 \\
        \frac{\sqrt{N}}{\Gamma_1^2} & \alpha = 0.5
    \end{cases}\, .
\end{align}
Unlike the 1D array, Dicke superradiance is always attainable in 2D arrays in both homogeneous media and the slab. However, in the slab case, the collective behavior can be achieved at farther separations, again confirming the effect of the prolonged far-field in mediating longer-range interactions.

\emph{Directional Dicke superradiance --} To assess the emergence of Dicke superradiance, we employ an equivalent criterion based on the instantaneous behavior of the photon emission rate at $t = 0$~\cite{Robicheaux2021}. This metric offers additional information by enabling the explicit analysis of directional information, such as the emitted photon statistics along a specific direction $\phi$ in the array plane. This approach allows for a deeper investigation of exotic features that may arise from the interplay between geometric arrangements and collective effects. 

The rate of emitted photons in the $\textbf{k}_f$ direction is determined as 

\begin{equation}
    \gamma(t,\textbf{k}_f) = \Gamma_0 \sum_n \left[\braket{\hat{\sigma}_{eg}^{(n)} \hat{\sigma}_{ge}^{(n)}}(t) + \sum_{m \ne n} e^{i\theta_{nm}} \braket{\hat{\sigma}_{eg}^{(m)} \hat{\sigma}_{ge}^{(n)}}(t)\right]\, ,
\end{equation}

where $\theta_{nm}$ is related to the relative phase of the exchanged photon between pairs $(m,n)$. 

The emergence of superradiance can be analyzed by examining the change in the emission rate at early times. Specifically, if $\dot{\gamma}(0,\textbf{k}_f) \geq 0$, the photon emission rate is enhanced due to the presence of other emitters, indicating the onset of superradiance. This criterion is calculated and visualized in the two-dimensional maps shown in Figs.~\ref{fig: 1D-Ypol}(b), (c), (e), and (f) for 1D arrays, and Figs.~\ref{fig: 1D-Ypol}(a)--(d) for 2D arrays.

\section{Z-polarized emitters}~\label{app: z-dipoles} 
This appendix provides complementary information about collective effects emerging in Z-polarized emitters.

\begin{figure*}
    \centering
    \includegraphics[width=0.78\linewidth]{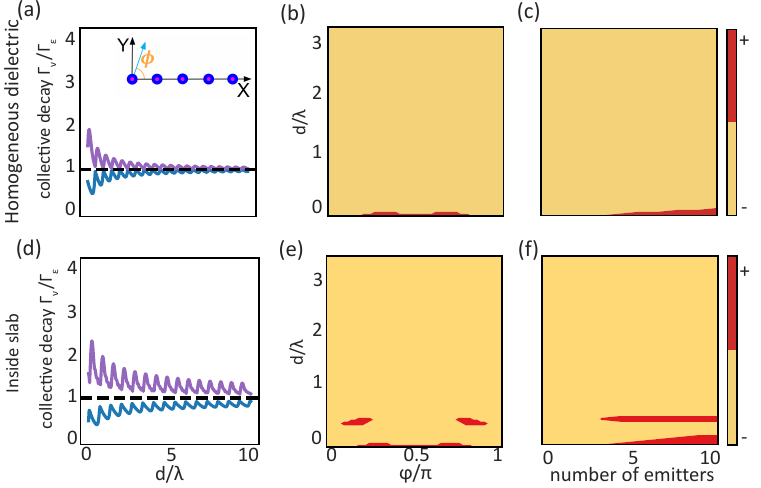}
    \caption{\textbf{Total and directional superradiance in a 1D array of Z-polarized emitters.} The collective decay rate of five emitters in a (a) homogeneous dielectric and (d) dielectric slab. The purple curve corresponds to the superradiant mode, while the blue curve corresponds to the subradiant one. Directional superradiance of a five-emitter 1D array in (b) a homogeneous dielectric and (e) dielectric slab, as a function of atomic separation ($d/\lambda$) and the in-plane angle $\phi$. The areas colored in red correspond to the regions where superradiance can be observed. The emergence of Dicke superradiance as a function of the number of emitters in a 1D array and their separations at $\phi=0.22\pi$ inside a (c) homogeneous dielectric medium and (f) slab. In all calculations $n = 3.5$ and $W = 200$ nm.}
    \label{fig: 1D-Zpol}
\end{figure*}

\begin{figure}
    \centering
    \includegraphics[width=1\linewidth]{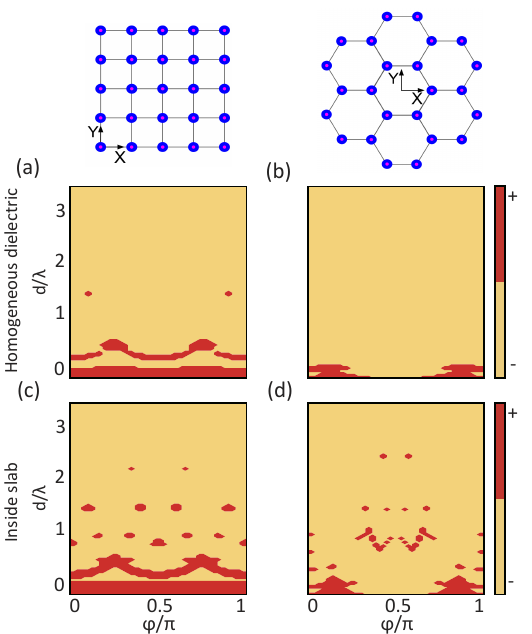}
    \caption{\textbf{Geometry effect on the total and directional superradiance in 2D arrays of Z-polarized emitters.} Directional superradiance for Z-polarized dipoles arranged in a 2D geometry inside a homogeneous dielectric ((a) and (b)) and a slab ((c) and (d)), delimited by red regions. Emitters in (a) and (c) are arranged on a square lattice with 25 sites, while the emitters in (b) and (d) are arranged on a hexagonal lattice with 24 sites, as depicted by small schematics on the first row. Regardless of the lattice geometry, the attainable directional superradiance regions are extended in a slab compared to a homogeneous medium. In all calculations $n = 3.5$ and $W = 200$ nm. Note that the hexagonal tiling will cover more area than the square one for the same lattice spacing.}
    \label{fig: 2D-Zpol}
\end{figure}

\begin{figure}
    \centering
    \includegraphics[width=0.97\linewidth]{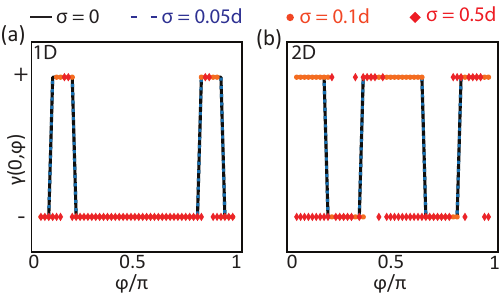}
    \caption{\textbf{Disorder effects on 1D and 2D Z-polarized emitters within a dielectric slab.} Directional superradiance 25 emitters in (a) 1D linear and (b) 2D square arrangement, with $d = 0.54 \lambda$. The black solid line, blue dashed line, orange dots, and red diamonds show the results for different disorders of $\sigma=0,~ 0.05d,~ 0.1d$, and $0.5d$, respectively.}
    \label{fig: Noise effects in Zpol}
\end{figure}

\end{document}